\documentclass[aps,twocolumn,prd,nofootinbib,showpacs]{revtex4}

\usepackage{amsmath}
\usepackage{graphicx}% \usepackage{dcolumn}
\usepackage{amssymb}

\newcommand{\form}[1]{\boldsymbol{#1}}
\newcommand{\order}[1]{\mathcal{O}\!\left(#1\right)}
\newcommand{\ud}{\mathrm{d}}
\newcommand{\ue}{\mathrm{e}}

\newcommand{\ur}{\mathrm{r}}

\newcommand{\us}{\mathrm{s}}

\newcommand{\uh}{\mathrm{h}}

\newcommand{\vh}{v_\uh}
\newcommand{\vs}{v_\us}
\newcommand{\vinf}{v_\infty}
\newcommand{\rhoh}{\rho_\uh}
\newcommand{\rhos}{\rho_\us}

\newcommand{\Ang}{\Theta}
\newcommand{\dsh}{d_{\mathrm{sh}}}
\newcommand{\const}{\mathrm{const.}}

\newcommand{\hfunc}{f_\uh}
\newcommand{\sfunc}{f_\us}

\begin{document}

\title{Charged seven-dimensional spacetimes with spherically symmetric
  extra-dimensions}

\author{Antonio De Felice} \email{antonio.defelice@uclouvain.be}
\affiliation{Theoretical and Mathematical Physics Group, Centre for
  Particle Physics and Phenomenology, Louvain University, 2 Chemin du
  Cyclotron, 1348 Louvain-la-Neuve (Belgium)}

\author{Christophe Ringeval}
\email{christophe.ringeval@uclouvain.be}
\affiliation{Theoretical and Mathematical Physics Group, Centre for
  Particle Physics and Phenomenology, Louvain University, 2 Chemin du
  Cyclotron, 1348 Louvain-la-Neuve (Belgium)}

\date{\today}

\begin{abstract}
  We derive exact solutions of the seven-dimensional Einstein-Maxwell
  equations for a spacetime exhibiting Poincar\'e invariance along
  four-dimensions and spherical symmetry in the extra-dimensions. Such
  topology generically arises in the context of braneworld models. Our
  solutions generalise previous results on Ricci-flat spacetimes
  admitting the two-sphere and are shown to include wormhole
  configurations. A regular coordinate system suitable to describe
  the whole spacetime is singled-out and we discuss the physical
  relevance of the derived solutions.
\end{abstract}
\pacs{04.50.+h, 11.10.Kk, 98.80.Cq}
\maketitle

%\begin{document}

\section{Introduction}

Extra-dimensions are considered to be a key ingredient for explaining
the quantum behaviour of fundamental theories. Starting from the work
of Kaluza and Klein~\cite{Nordstrom:1914,Kaluza:1921,Klein:1926}), the
embedding of our universe into an higher-dimensional space has been
invoked to give an explanation of apparently unrelated
four-dimensional phenomena. The unification of forces at first, but
also more recently the hierarchy problem, the acceleration of the
universe are different issues which have been addressed by using the
features of the extra-dimensions.

It is therefore of interest to look how gravity behaves in a
higher-dimensional world, that is to find exact non-perturbative
solutions for the classical equations of motion. Many works have been
devoted to this task, especially in the framework of Superstring
theories in which supersymmetry and compactification play a major
role~\cite{Polchinski:1998rq}. Various black brane configurations have
been studied in the literature as exact solutions of Superstring low
energy effective actions. For this reason, most of them still exhibit
some amount of Supersymmetry, as well as non-trivial configuration of
higher-dimensional form fields~\cite{Horowitz:1991cd, Gueven:1992hh,
  Gauntlett:1992nn, Duff:1993ye, Liu:1999ai, Stelle:1998xg}. In this
paper, motivated by Cosmology in presence of extra-dimensions, we
adopt a (very) low energy effective approach and remain in the
framework of standard General Relativity. In this context, it is clear
that the simplest braneworld model one can think of consists of a
four-dimensional Minkowski manifold embedded in a higher dimensional
spacetime (bulk). For a five-dimensional anti de-Sitter bulk, one
would recover Randall--Sundrum (RS)
constructions~\cite{Randall:1999ee,Randall:1999vf} whereas
asymptotically flat extra-dimensions would be reminiscent with the
Dvali--Gabadadze--Porrati (DGP) braneworld models~\cite{Dvali:2000hr,
  Dvali:2000xg}. {}From a classical Field Theory approach, it has
recently been shown in Ref.~\cite{DeFelice:2008af} that the DGP
gravity confinement mechanism along a four-dimensional world-volume
can be realised in the core of a 't~Hooft--Polyakov seven-dimensional
hypermonopole. In fact, seven dimensions is the minimal number of
spacetime dimensions for which the trapping of gravitons by curvature
effects may occur, and this is closely related to the existence of a
foliation of the extra-dimensions by positively curved
hypersurfaces~\cite{Ringeval:2004ju}. Although the full system of
Einstein--Yang--Mills equations have been numerically solved in
Ref.~\cite{DeFelice:2008af}, under the above-mentioned symmetry, one
may wonder if some exact solutions could not be derived. In fact, far
from the core of a 't~Hooft--Polyakov hypermonopole, that is to say
where the $SO(3)$ Higgs and gauge fields reach their vacuum
expectation values, the remaining unbroken gauge symmetry is
$U(1)$. As a result, at large distances, the stress-tensor content of
the theory is similar to a higher dimensional Dirac
monopole~\cite{Nepomechie:1984wu}. Motivated by this picture, we
derive in this paper exact analytical solutions of the
Einstein-Maxwell equations and study the resulting curved spacetime in
the presence of an electrical/magnetic field seeded by an $U(1)$
two-form. In the braneworld framework, we are looking for static
Einstein--Maxwell solutions for a compactification of a
seven-dimensional spacetime into $M^4\times R\times S^2$. The
solutions studied can therefore be viewed as the spatially extended
generalisation of the Dirac monopole in four dimensions, where a gauge
field, carrying a magnetic charge $1/q$, introduces a non-vacuum
structure for the spacetime. Notice that since Poincar\'e invariance
is not broken along the brane, such a topology differs from
higher-dimensional black hole solutions on the
brane~\cite{Chamblin:1999by, Dadhich:2000am, Gibbons:2002bh,
  McFadden:2004ni, Galfard:2005va, Gibbons:2008gg}. However the
electrically dual action would involve a dual five-form field which is
reminiscent with higher-dimensional generalisation of
Reissner--Nordstr\"om solutions~\cite{Beker:2007}. More specifically,
such extra-dimensional topology have been studied for the vacuum case
in five-dimensions in Refs.~\cite{Gross:1983hb, Davidson:1985zf} and
recently revisited in Refs~\cite{PoncedeLeon:2005jp, Millward:2006gg,
  Lake:2006hs}. Apart from the different number of dimensions, our
approach generalises these results for non-vacuum spacetimes. The
electric dual counterpart of our configuration has been discussed in
Refs.~\cite{Bronnikov:1996df, Hassaine:2007py} for even spacetime
dimensions only and spherically symmetric in all spatial
dimensions. As discussed in the following, we will encounter the same
kind of problem of apparently ``truncated'' solutions described in
Ref.~\cite{Lake:2006hs}, where the coordinates were not able to
explore the whole manifold. This can be solved with a choice of a
suitable coordinate system that we describe in the following. Although
not using these coordinates is convenient for the study of exterior
spacetimes, as the one discussed in Ref.~\cite{DeFelice:2008af}, this
could lead to misinterpreting the physical content of the theory. In
our regular coordinate system, we show, in particular, that some
solutions of the Einstein-Maxwell system simply correspond to charged
wormholes configurations~\cite{Kanti:2002fx, Rogatko:2003kj,
  Rogatko:2006gg}. Such exact solutions, and especially the treatment
of the coordinates introduced here, may shed some light into other
similar cases.

In Sect.~\ref{sec:model} we introduce the model. In
Sect.~\ref{sec:vacuum} we rederive the vacuum solutions within our
coordinate system and in seven dimensions. We moreover recap the
merging of different and apparently unrelated patches. In
Section~\ref{sec:charged}, we move on to the general solution we are
interested in, namely in the presence of a magnetic field in the
extra-dimensions. We then conclude in the last section.

\section{The model}
\label{sec:model}

The model we consider is defined by the standard Einstein--Maxwell
action
\begin{equation}
\label{eq:action_pole}
S = \int \sqrt{-g}\, \ud^7 x \left(\frac{R}{2\kappa^2} -\dfrac{1}{4}
  F_{AB} F^{AB} \right) ,
\end{equation}
where the field strength 2-form $\form{F}$ can be defined in terms
of a 1-form $\form{A}$, as $\form{F}=\form{\ud A}$, so that the
Lagrangian is invariant under a $U(1)$ gauge transformation
$\form{A}\to\form{A}+\form{d}\chi$. Capital Latin indices run
from 0 to 6, whereas Greek indices from 0 to 3. The dimensions of the
gauge field are $[C^{aM}]= M^{5/2}$, those of the seven dimensional
Newton constant are $[\kappa^2]=M^{-5}$, and for the an electric
charge $[q]=M^{-3/2}$. The stress-energy tensor is therefore
\begin{equation}
T^A{}_{B}=F^{AN} F_{BN}-\dfrac14 F^2 \delta^A{}_{B} .
\end{equation}
Finally the Einstein equations are
\begin{equation}
G^A{}_B=\kappa^2\,T^A{}_B ,
\end{equation}
while the Maxwell equations for the 2-form read as $\form{\ud F}=0$, or
\begin{equation}
F_{AB,C}+F_{BC,A}+F_{CA,B}=0 ,
\end{equation}
and $\nabla_B F^{AB}=0$.

\subsection{Background metric}

We are looking for background solutions with spherical symmetry in the
extra-dimensions, i.e., a compactification into $M^4\times R\times
S^2$. Therefore, we impose the following ansatz for the metric
\begin{equation}
\label{eq:metric}
\ud s^2=A(r)\left(-\ud t^2+ \ud \vec x^{\,2} \right)+\xi(r) \ud
r^2+r^2 \ud \Omega^2,
\end{equation}
with
\begin{equation}
\ud \Omega^2 = \ud\theta^2+\sin^2 \theta \ud\phi^2.
\end{equation}
Furthermore the 2-form field is assumed to be of the form
\begin{equation}
\form{F}=f(r,\theta,\phi) \form{\ud}\theta\wedge\form{\ud}\phi.
\end{equation}
The Maxwell equations, $\form{\ud F}=0$, immediately imply that
$\partial f/\partial r=0$, or $f=f(\theta,\phi)$. Then we have
\begin{align}
\nabla_AF_\theta{}^A{}&=\frac1{r^2\sin^2\theta}\frac{\partial f}
      {\partial\phi}=0 ,\\ \nabla_AF_\phi{}^A&=-\frac
      1{r^2\sin\theta} \left[\frac{\partial
      f}{\partial\theta}\sin\theta-f\cos\theta\right]=0 ,
\end{align}
which have the non-trivial solution
\begin{equation}
f=-\frac{\sin\theta}q ,
\end{equation}
and where the integration constant $q$ is the electric charge.

This ansatz automatically solves the Maxwell equations for the given
metric, independently of the profiles for the fields $A(r)$ and
$\xi(r)$. In the three extra-dimensions, it is now possible to define
a three-vector, with components $B^i$, which is the magnetic field
associated to the 2-form. With
\begin{equation}
B^i=-\varepsilon^{ijk}F_{jk}/\sqrt{{}^{(3)}g},
\end{equation}
where $i,j,k\in\{4,5,6\}$, one gets
\begin{equation}
B_r=\dfrac{1}{qr^2}\,.
\end{equation}
Therefore $1/q$ appears as a magnetic charge, the magnetic field
is purely radial and we are in presence of a spatially extended Dirac
magnetic monopole at the origin of the coordinate system. The
associated stress tensor is non-vanishing and reads
\begin{equation}
T^0{}_0=T^x{}_x=T^r{}_r=-T^\theta{}_\theta=-T^\phi{}_\phi=-\frac{1}{2 q^2\,r^4} .
\end{equation}
In fact, this ansatz for the 2-form is the only one consistent with
the choice of the metric ansatz, as the other components for $F_{AB}$
have no solutions for the Einstein equations.

\subsection{Equations of motion}

{}From the above ansatz, the Einstein--Maxwell equations read
\begin{align}
\dfrac32 {\frac { {A'}^{2}}{A^{2}\xi}} +{
\frac {4 A'}{A \xi  r}}-\frac1{r^2}+{\frac {1}{\xi  {r}^{2}
}}+\frac{\kappa^2}{2q^2r^4} &=0 ,
\label{eqz1}\\
{\frac {2 A'}{A \xi r}}
 -\dfrac12 {\frac {\xi'  }{ \xi^{2} r}}
+\dfrac12 {\frac {{A'}^{2}}{A^{2} \xi }}
+{\frac {2 A''}{A \xi}}
-{\frac { A' \xi'}{\xi^{2}A}}-\frac{\kappa^2}{2q^2r^4} & =0,
\label{eqz3}\\
\dfrac32 {\frac {A'' }{A  \xi}}
 -\dfrac34 {\frac { A'\xi'}{\xi^{2}A}}
+{\frac {3 A'}{A  \xi r}}
-{\frac {\xi'}{\xi^{2}r}}
-\frac1{r^2}+{\frac {1}{\xi
    {r}^{2}}}+\frac{\kappa^2}{2q^2 r^4} &=0,
\label{eqz2}
\end{align}
where the prime denotes derivative with respect to $r$. {}From
Eq.~(\ref{eqz3}) it is possible to solve for $A''$. One can then
eliminate $A''$ from Eq.~(\ref{eqz2}) and solve for ${A'}^2$ in terms
of $A'$, $\xi'$, $A$ and $\xi$. Then, using this, one can eliminate
${A'}^2$ from Eq.~(\ref{eqz1}). Finally, one gets
\begin{equation}
  \frac{A'}A=\dfrac14{\frac {\xi' }{
      \xi}}-\dfrac25{\frac {\xi {\kappa}^{2}}{{r}^{3}{q}^{2}}}
  +\dfrac12{\frac {\xi-1 }{r}} \,.
\label{eqAxi}
\end{equation}
One can use this equation for $A$, together with Eq.~(\ref{eqz1}) to
obtain
\begin{equation}
\label{eq:odexir}
  \xi'=-{\frac {2 \xi^{2}}{r}}-\dfrac{10}3{
    \frac {\xi }{r}}+{\dfrac {16}{5}}{\frac {\xi ^{2}{r_0}^{2}}{{r}^{3}}}
  \pm\dfrac43{\frac {
      \xi\sqrt {(6\xi+10)  {r}^{2}
        -6{r_0}^{2}\xi  }}{{r}^{2}}}\,,
\end{equation}
where $r_0$ stands for
\begin{equation}
r_0\equiv \dfrac{\kappa}{\sqrt{2} q}\,.
\end{equation}
The vacuum case is obtained in the limit $r_0 \to 0$. If $r_0\neq0$ we
can define the dimensionless variable $\rho=r/r_0$ and the
previous equation becomes
\begin{equation}
\xi'=-{\frac {2 \xi^{2}}{\rho}}-\dfrac{10}3{
\frac {\xi }{\rho}}+{\dfrac {16}{5}}{\frac {\xi ^{2}}{{\rho}^{3}}}
\pm\dfrac43{\frac {
\xi\sqrt {(6\xi+10)  {\rho}^{2}
-6\xi  }}{{\rho}^{2}}} \,,
\label{dixi}
\end{equation}
where now a prime denotes differentiation with respect to $\rho$, and
one can see that there are two branches. In order to study the
presence of singularities, it is convenient to study both the Ricci
curvature $R$, and the Kretschmann scalar defined as
\begin{equation}
K=\dfrac14R_{ABCD}R^{ABCD}.
\end{equation}
This second scalar invariant will be especially useful for the vacuum
case, when $R$ vanishes on the solutions of the equations of
motion. Their expression for the metric~(\ref{eq:metric}) are given in
the Appendix~\ref{app:grinv}.

One can use Eq.~(\ref{eqAxi}) in order to express $R$ and $K$ only in
terms of $\xi$ and its derivatives. Moreover, for the solutions of the
Einstein equations one finds that $R\propto T$, where $T$ is the trace
of the stress tensor. As a result, for the non-vacuum case, one
has on-shell
\begin{equation}
  r_0^2R=\dfrac{6}{5\rho^{4}}\, ,
\end{equation}
which states that only for $\rho=0$ the Ricci scalar blows up. Before
facing the problem of solving Eq.~(\ref{dixi}), it is worthy
illustrating the method and the choice of a suitable coordinate system
for the vacuum solutions. This will allow to make contact with
previous works, albeit in a different coordinate
system~\cite{Gross:1983hb, Davidson:1985zf, PoncedeLeon:2005jp,
  Millward:2006gg, Lake:2006hs}.

\section{Vacuum solutions}
\label{sec:vacuum}

The vacuum equations are obtained in the limit $q\to\infty$ (or
$r_0\to0$). Defining in this section $\rho \equiv r$,
Eq.~(\ref{eqAxi}) simplifies to
\begin{equation}
\frac{A'}A=\dfrac { \xi' \rho+2\xi^2-2\xi  }{4\rho\xi}\, ,
\label{Av}
\end{equation}
which can be integrated into
\begin{equation}
  A(\rho)=\exp \left(\int^\rho\frac { \xi' r+2\xi^2-2\xi
    }{4r\xi} \ud r \right).
\label{Asol}
\end{equation}
This expression can be further simplified into
\begin{equation}
\label{eq:Adiff}
  \frac{ A\sqrt{\rho}}{\xi^{1/4}} = \exp\left(\dfrac{1}{2}
  \int^\rho\frac{\xi(r)}{r} \ud r\right),
\end{equation}
Finally, as for $\xi$, Eq.~(\ref{eq:odexir}) reads
\begin{equation}
\rho\frac{\xi'}\xi=-\dfrac{10}3-2\xi\pm\dfrac43\sqrt{10+6\xi} \,.
\label{dixiv}
\end{equation}
The plus and minus sign indicate that there are two branches for the
$\xi(\rho)$ solutions. This equation can be integrated by a separation
of variables. The above expression requires $\xi\geq- 5/3$, and the
limiting case $\xi=-5/3$ is a particular solution of
Eq.~(\ref{dixiv}), for which $A=A_0\rho^{-4/3}$. This solution
corresponding to a timelike $\rho$ coordinate will not be considered
as a physical one. Notice that switching the sign of $A_0$ does not
cure the problem, as the three coordinates, $x,y,z$, would all become
timelike.

\subsection{First branch}

\begin{figure}[t]
\begin{center}
\includegraphics[width=8.5cm]{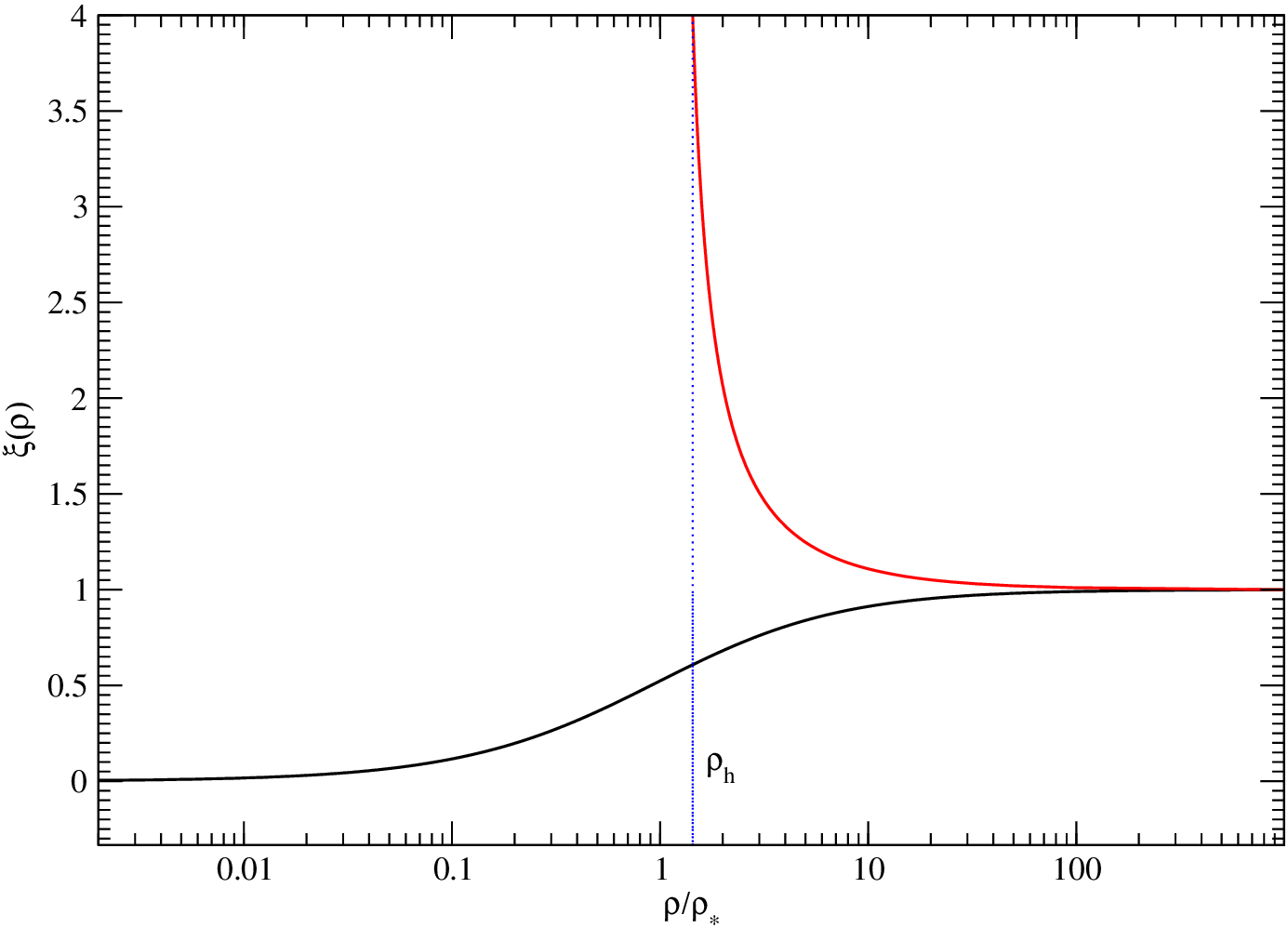}
\caption{Metric coefficient $\xi(\rho)$ for the first branch ``$+$''
  of the vacuum solution. There are two solutions, one exhibiting a
  null surface at $\rhoh = \rho/\rho_*=3^{\sqrt{2/5}-1/2}$ and the
  other a naked singularity at $\rhos=0$. In this coordinate system,
  the solution exhibiting the null surface is incomplete.}
\label{fig:vac1}
\end{center}
\end{figure}

After separating variables in Eq.~(\ref{dixiv}), and choosing the plus
sign for the square root, one gets
\begin{equation}
  \int_{}^\xi
  \frac{\ud \tilde\xi}{\tilde\xi\left( - \dfrac{10}3-2\tilde\xi +
    \dfrac43 \sqrt{10+6\tilde\xi } \right)}
  =\ln\rho ,
\end{equation}
whose solution reads
\begin{equation}
\label{eq:xivacplus}
  \frac{\left\vert\sqrt{3 \xi
        +5} - \sqrt{5}\right\vert^{2 \sqrt{2/5}}}{
    |\xi| ^{\sqrt{2/5}-1/2}|\sqrt{3 \xi +5}-2\sqrt{2}|}=\dfrac{\rho}{\rho_*}\, ,
\end{equation}
where $\rho_*>0$ is an integration constant. The solution is defined
only on three intervals, $-5/3<\xi<0$, $0<\xi<1$, and $\xi>1$. Among
them, the first one again corresponds to a timelike spatial
coordinates and will not be considered as a physical one.

As can be seen in Eq.~(\ref{eq:xivacplus}), $\xi(\rho)\to \infty$ for
\begin{equation}
  \rho = \rhoh =3^{\sqrt{2/5}-1/2}\rho_*.
\end{equation}
Since the Kretschmann scalar given in Eq.~(\ref{eq:Kscalar}) remains
finite at that point, we are in presence of a coordinate singularity
only. In fact, as can be checked from Eq.~(\ref{eq:metric}), since
$\xi(\rho)$ is divergent at that point, the hypersurface $\rho=\rhoh$
is actually a null surface. Notice however that the redshift
counterpart $A(\rhoh)$ remains finite. As we discuss in
Sect.~\ref{sect:wh}, this null surface is not an horizon. On the
other hand, there is a singularity, i.e.\ $K\to\infty$, at the point
where $\xi\to0^+$:
\begin{equation}
  \rho= \rhos = \rho_{*}\lim_{\xi\to0^+}\frac{\left(\sqrt{3 \xi
        +5}-\sqrt{5}\right)^{2 \sqrt{2/5}}}{
    \xi^{\sqrt{2/5}-1/2}\left(2\sqrt{2}-\sqrt{3 \xi +5}\right)}=0,
\end{equation}
which is the origin of the coordinate system. Finally, for non-compact
spacetime, the coordinate $\rho$ should reach $\rho\to+\infty$. This
is possible if $\xi\to1^+$ or if $\xi \to 1^-$. As a result, the
spacetime is necessarily asymptotically Minkowski. In
Fig.~(\ref{fig:vac1}), we have plotted the metric coefficient
$\xi(\rho)$ given by Eq.~(\ref{eq:xivacplus}). Notice that we find two
disjoint solutions, even so we are here considering only the ``$+$''
branch. As discussed in Ref.~\cite{Lake:2006hs}, it is not clear what
happens for $\rho<\rhoh $ since the $\rho$ coordinates
are not able to describe the whole spacetime, leaving an apparently
empty spot in the manifold (see Fig.~\ref{fig:vac1}).

\subsection{Second branch}
\begin{figure}
\begin{center}
\includegraphics[width=8.5cm]{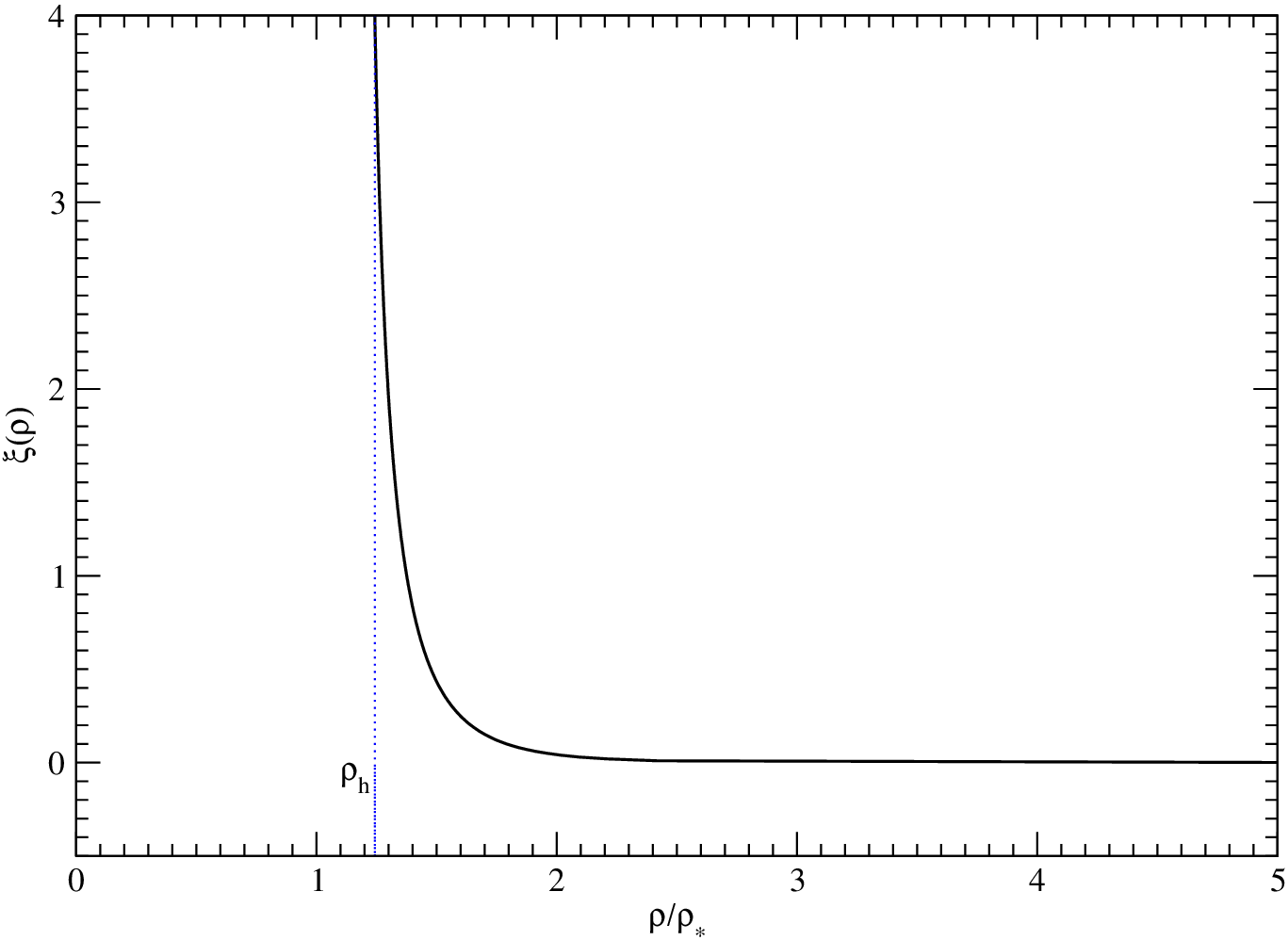}
\caption{Metric coefficient $\xi(\rho)$ for the ``$-$''-branch of the
  vacuum solution. The solution has a null surface at
  $\rho/\rho_*=3^{\sqrt{2/5}-1/2}$, and a naked singularity as $\rho
  \to \infty$. In this coordinate system, the solution is incomplete.}
\label{fig:vac:2}
\end{center}
\end{figure}

Similarly to the previous discussing, but choosing now the minus sign
for the square root in Eq.~(\ref{dixi}), one gets with $\xi>0$,
\begin{equation}
\label{eq:xivacneg}
  \frac{\left(\sqrt{3 \xi
        +5}+\sqrt{5}\right)^{2 \sqrt{2/5}}}{
    \xi ^{\sqrt{2/5}-1/2} \left(\sqrt{3 \xi
        +5}+2\sqrt{2}\right)} = \dfrac{\rho}{\rho_*}\, .
\end{equation}
The position of the null surface is again given by the point at which
$\xi\to+\infty$, i.e., for
\begin{equation}
  \rhoh = 3^{\sqrt{2/5}-1/2}\rho_* .
\end{equation}
Asymptotically, we want $\rho \to \infty$ which is obtained for
vanishing denominator in Eq.~(\ref{eq:xivacneg}), i.e., for $\xi \to
0^+$. Contrary to the ``$+$''-branch, there is therefore only one
solution for the ``$-$''-branch. Notice that the point for which
$\xi\to0^+$ is also a singularity since $K\to\infty$. As a result, this
solution has both an horizon at a finite value of $\rho$ and a
singularity for $\rho\to\infty$. The corresponding solution has been
plotted in Fig.~(\ref{fig:vac:2}). Once more, the situation for
$\rho<\rhoh$ is unclear. As we shall see in the next
section, this issue is only due to a bad choice of the coordinate
system.

\subsection{Filling the gaps}

The variable $\rho$ does not seem to clarify the situation for the
whole space of solutions. Some solutions look truncated at finite
distance, other seem to describe different behaviours for the same
$\rho$. It is possible that the different patches may be joined
through a different and more suitable choice of coordinates. Mostly
the behaviour of $\xi$ is quite unclear for $\rho<\rhoh$. Pursuing the
idea that all this situation is due to a bad choice of coordinates,
we introduce a new coordinate system joining the two branches in
the next sections.

\subsubsection{First branch}

Let us introduce the new coordinate $v(\rho)$ such that
\begin{equation}
  \xi(\rho)=\frac53 \left( u^2-1\right) ,
\end{equation}
where
\begin{equation}
u=\frac{5\gamma_0-2 \sqrt{10}
        \rho v}{5\gamma_0 -5 \rho v}
\quad\textrm{and}\quad
\gamma_0\equiv\frac{13+4\sqrt{10}}{3}\,.
\end{equation}
In terms of $v$, the equation of motion~(\ref{dixi}) becomes
\begin{equation}
  \dfrac{\ud v}{\ud \rho} =\frac{v}{\rho\left[v(\rho) \rho-\gamma_0\right]} \,.
\end{equation}
This equation can be inverted to obtain $\rho(v)$:
\begin{equation}
  \dfrac{\ud \rho}{\ud v} = \rho^2-\gamma_0\frac{\rho}{v}\, ,
\label{eq1br}
\end{equation}
This is a Riccati equation, which can be linearised with another
change of variable. Introducing $s(v)$ such that
\begin{equation}
\rho(v) = - \dfrac{1}{s} \dfrac{\ud s}{\ud v} \,,
\end{equation}
one can rewrite Eq.~(\ref{eq1br}) as
\begin{equation}
\dfrac{\ud^2 s}{\ud v^2} + \dfrac{\gamma_0}{v} \dfrac{\ud s}{\ud v} = 0 .
\end{equation}
This equation has the symmetry $v\to-v$, and the general solution is
\begin{equation}
  s=c_1|v|^{1-\gamma_0}+c_2 .
\end{equation}
Therefore for every solution $s(v)$ with $v>0$ there is another
solution $s(-v)=s(v)$. Without loss of generality, we can study only
the region $v>0$. It is interesting to notice that under this symmetry
$\rho(-v)=-\rho(v)$, and $\xi(-v)=\xi(v)$. Finally, the general
solution for $\rho$ reads
\begin{equation}
  \rho_1(v)=\frac{\gamma_0-1}{v\left(1-C|v|^{\gamma_0-1}\right)} \,,
\end{equation}
where $C$ is a dimensionful integration constant. It is clear that
$\rho_1$ does not seem to be necessarily positive for all value of $v$
and $C$.

\subsubsection{Second branch}

As for the first branch, the new coordinate $v(\rho)$ is defined by
\begin{equation}
  \xi(\rho) =\frac53\left(u^2
    -1\right) ,
\end{equation}
where
\begin{equation}
  \label{eq:u2vc}
  u=-\frac{5\gamma_0+2\sqrt{10}\rho v
      }{ 5\gamma_0 +
        5\rho v} ,
\end{equation}
and the differential equation for $v$ becomes
\begin{equation}
  \dfrac{\ud v}{\ud \rho} + \frac{v}{\rho \left[v(\rho) \rho-\gamma_0\right]
  }
  = 0,
\end{equation}
and inverted into
\begin{equation}
\dfrac{\ud \rho}{\ud v} = -\rho^2-\gamma_0\frac{\rho}{v}\, .
\label{eq2br}
\end{equation}
Denoting by $\rho_2(v)$ the solution of the above equation
and comparing Eqs.~(\ref{eq1br}) and (\ref{eq2br}), one finds
$\rho_2(v)=-\rho_1(v)$. Therefore, the solution reads
\begin{equation}
  \rho_2(v)= -\frac{\gamma_0-1}{v \left(1-C|v|^{\gamma_0-1} \right)} \,.
\end{equation}

\subsubsection{Joining the branches}

For any value of $C$ and $v$, either $\rho_1(v)$ or $\rho_2(v)$ will
be positive, as $\rho_2=-\rho_1$. Hence, for all $v\ne0$, we can set
\begin{equation}
\rho(v) \equiv |\rho_1(v)|=|\rho_2(v)|.
\end{equation}
Similarly, for the two branches, the metric factor verifies
$\xi_1(v)=\xi_2(v)$ for all $v\ne 0$ and we define
\begin{equation}
  \xi(v) \equiv \xi_1(v)=\xi_2(v).
\end{equation}

In terms of the $v$ coordinate, the two branches are therefore unified
and, assuming from now that $v>0$, we have
\begin{equation}
\label{eq:rhoxivac}
\begin{aligned}
   \rho(v) &=\dfrac{\gamma_0-1}{\left|v-C v^{\gamma_0} \right|} ,\\
   \xi(v) & = \dfrac{20C \gamma_0 v^{\gamma_0+1}}{3 \left[C\gamma_0
       v^{\gamma_0}-v\right]^2} \,.
\end{aligned}
\end{equation}

\subsubsection{Singularities}

For all $C\ne0$ these solutions give
\begin{equation}
\lim_{v\to+\infty}\rho=0 ,\qquad \lim_{v\to+\infty}\xi=0 ,
\end{equation}
and, as can be checked with the Kretschmann scalar, there is a
singularity for $v\to \infty$ ($\rho \to 0$). For vanishing $v$, we
recover the asymptotic singularity
\begin{equation}
  \lim_{v\to0}\rho=+\infty ,\qquad \lim_{v\to0}\xi=0.
\end{equation}

For intermediate values of $v$, we recover the null surface at
the values of $v$ such that $\xi\to\infty$. As can be checked in
Eq.~(\ref{eq:rhoxivac}), the existence of coordinate singularities in
the intermediate region depends on the values of the integration
constant $C$. Denoting by $\vh$ the location of this surface
Eq.~(\ref{eq:rhoxivac}) yields
\begin{equation}
\label{eq:vxiinf}
\vh = \dfrac{1}{\left(\gamma_0 C\right)^{1/(\gamma_0-1)}}\, ,
\end{equation}
provided $C>0$, and no solution if $C<0$.

We can also separate the different regions of spacetime by looking at
the behaviour of $\rho(v)$ for finite values of $v$. Denoting by
$\vinf$ the point at with $\rho\to\infty$, from
Eq.~(\ref{eq:rhoxivac}), one gets
\begin{equation}
\label{eq:vrhoinf}
\vinf = \dfrac{1}{C^{1/(\gamma_0-1)}}.
\end{equation} 
Also in this case, there is a solution only for $C>0$. Moreover, for
any non-vanishing values of $\vinf$ one has $\xi\to1$ and the
spacetime is asymptotically flat.

\begin{figure}
\begin{center}
\includegraphics[width=8.5cm]{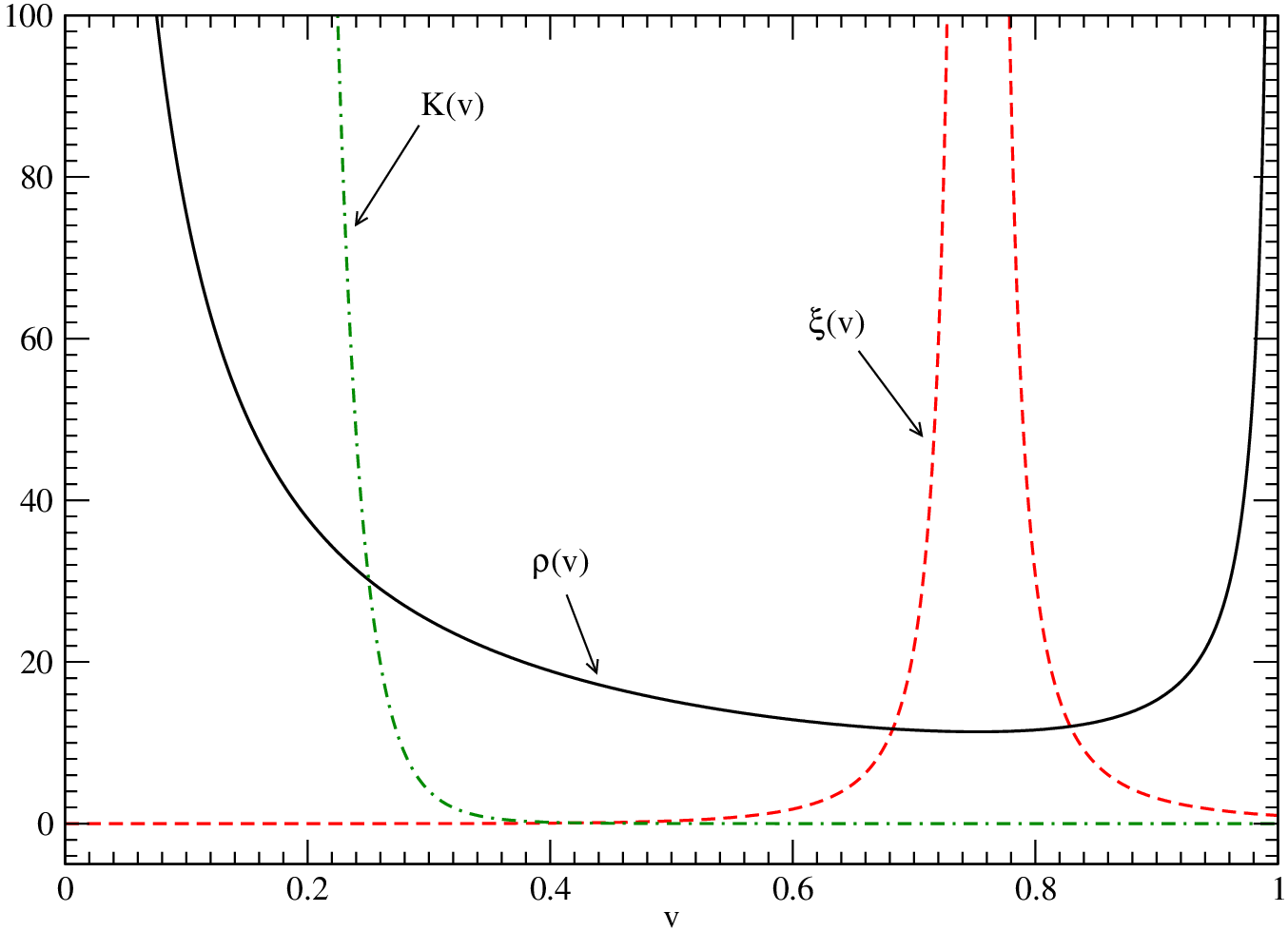}
\caption{Wormhole patch: a singularity is present at $v\to 0$,
  together with a null surface at $\rho=\rhoh$ and an asymptotically flat
  region for $v\to \vinf=1$. The behaviour of $\rho(v)$ is multivalued
  and describes a wormhole-like configuration. Notice that the
  divergence of $\xi$ is simply the results of the ill-defined
  $\rho$-coordinate system. The null surface is not an horizon and
  corresponds to the wormhole throat. Although the wormhole has no
  horizon, it exhibits a singularity on the brane side and at a finite
  proper distance from the throat. It is therefore not properly
  traversible.}
\label{fig:vac:ptc6}
\end{center}
\end{figure}

\begin{figure}[ht]
\begin{center}
  \includegraphics[width=8.5cm]{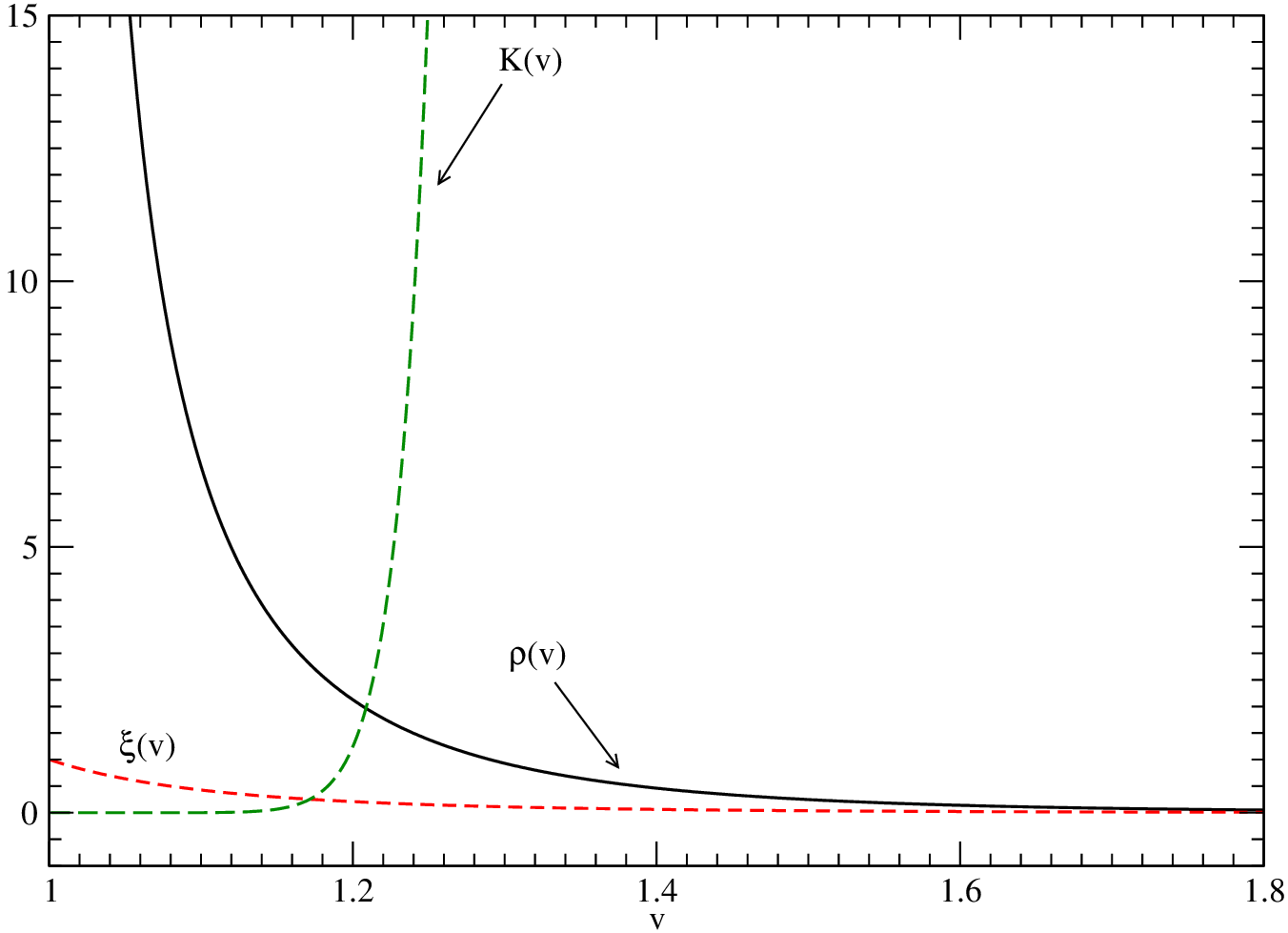}
  \caption{Naked singularity patch: an asymptotically flat region
    is reached as $v \to \vinf=1$ but there is a naked singularity as $\xi \to0$.}
\label{fig:vac:ptc7}
\end{center}
\end{figure}

{}From Eqs.~(\ref{eq:vxiinf}) and (\ref{eq:vrhoinf}), it is clear that
we always have $\vh < \vinf$. As a result, for positive $C$,
there are two different patches for the spacetime

\begin{itemize}

\item $0<v<\vinf$. A singularity is present for $v\to0$, together with
  a null surface at $\vh$, and an asymptotically flat region for $v
  \to \vinf$. We refer to this patch as the wormhole solution (see
  Fig.~\ref{fig:vac:ptc6}). Notice the multivalued and unbounded
  behaviour of $\rho(v)$ which is reminiscent with a wormhole
  solution. As we show in the next section, the divergence of
  $\xi(\rho)$ comes from the bad choice of the coordinate system: the
  null surface is not an horizon and traces instead a wormhole
  throat. This wormhole however connects the asymptotically flat
  region to a singularity, which is at a finite proper distance from
  the throat.

\item $v>\vinf$. The asymptotically flat region occurs for $v\to
  \vinf$ and there is a singularity for $v\to \infty$ (see
  Fig.~\ref{fig:vac:ptc7}). There is no horizon, neither wormhole
  configuration in that patch and the singularity is naked.
\end{itemize}

For completeness, let us consider the case $C<0$. {}From
Eq.~(\ref{eq:rhoxivac}), two singularities are now present for $v\to
0$ and $v\to \infty$, without any horizon and with $\xi<0$. This
solution exhibits two naked singularities and two timelike
coordinates. Therefore it is pathological and will not be discussed
any longer in the following.

\subsection{Wormhole solution}

\label{sect:wh}

All solutions discussed in the previous section exhibit singularities
and we discuss in this section the ``wormhole patch.'' As can be seen
in Fig.~\ref{fig:vac:ptc6}, it is a non-trivial merging of the first
and second branch of the $\rho$-coordinate system.

\subsubsection{Nature of the null surface}

Clearly, the $v$-coordinate system is non-singular for this solution
(except at most the physical singularity where $\xi\to0$), and the
metric becomes
\begin{equation}
\label{eq:vmetric}
\begin{aligned}
  \ud s^2 &= A\left[\rho(v)\right] (-\ud t^2 + \ud\vec x^2) + \xi(v)
  \left(\dfrac{\ud \rho}{\ud v}\right)^2 \ud v^2  \\
  & + \left[\rho(v)\right]^2 \ud \Omega^2 .
\end{aligned}
\end{equation}

{}From Eq.~(\ref{eq:rhoxivac}), one can express the metric coefficient
$g_{vv}$ as
\begin{equation}
\label{eq:gvv}
  g_{vv} = \xi(v) \left(\dfrac{\ud \rho}{\ud v}\right)^2 =
  \frac{20  \gamma_0(\gamma_0-1)^2 C v^{\gamma_0-1}}{3 \left(C  v^{\gamma_0
      }-v\right)^4} \,.
\end{equation}
This quantity does not blow up at $\vh$, namely for the points at
which $\xi\to\infty$, so that in these variables the coordinate
singularities are no longer present. In order to determine the
behaviour of $A(v)$, we still need to integrate Eq.~(\ref{eq:Adiff})
in terms of $v$, and in particular the integral
\begin{equation}
  I_1\equiv \dfrac{1}{2} \int \dfrac{\xi(v)}{\rho(v)} \dfrac{\ud \rho}{\ud v} \ud v.
\end{equation}
Explicitly, $I_1$ reads
\begin{equation}
  I_1 =- \dfrac{10}{3C} \int \dfrac{v^{\gamma_0-2} \ud v}
  { [v^{\gamma_0-1} - C^{-1}] [v^{\gamma_0-1} -(C\gamma_0)^{-1}]} \, ,
\end{equation}
which can be integrated exactly into
\begin{equation}
  I_1 = -\dfrac{1}{2} \ln \left | \dfrac{C v^{\gamma_0}-v}
    {C\gamma_0v^{\gamma_0}-v}\right | + \const \,,
\end{equation}
Therefore $A(v)$ simplifies to
\begin{equation}
\label{eq:Aofvinf}
A = A_0 |C|^{1/4} v^{(\gamma_0+1)/4} ,
\end{equation}
where $A_0^2=|C|^{1/(\gamma_0-1)}$ is a constant of integration fixed
according to the clock of an asymptotically flat observer. The
``redshift function'' $A(v)$ is therefore regular at the $v=\vh$
location, and, at this point, it does not vanish.

In this coordinate system, one can immediately check that the constant
radial hypersurfaces still correspond to constant $v$
hypersurfaces. From Eq.~(\ref{eq:vmetric}), it is then clear that the
lightlike geodesics can intercept and cross the null hypersurface
located along $\rho=\rho_h$. This hypersurface is therefore not an
horizon.

The nature of the coordinate singularity at $\rho=\rhoh$ can be
further clarified by studying the convergence $\theta_\ur$ of a
congruence of radial null geodesics. In fact, although the $v$
coordinates cure the coordinate singularity at the null hypersurface
location, it is not well defined asymptotically and one may question
the existence of an asymptotically flat observer at
infinity. Therefore, let us define a new radial variable as
\begin{equation}
  \label{eq:wm1}
  \zeta=\int_0^v\sqrt{\frac{g_{vv}(v')}{A(v')}}\,\ud v' ,
\end{equation}
so that the singularity is located at $\zeta=0$. Then, it is
straightforward to find a null coordinate system which is regular
everywhere and asymptotically flat by defining
\begin{equation}
\begin{aligned}
u &= t-\zeta\\
w &= t+\zeta .
\end{aligned}
\end{equation}
In this coordinate system, one gets
\begin{equation}
  \label{eq:wm2}
  \ud s^2=-A[v(u,w)]\ud u\,\ud w+A\,\ud\vec x^2+\rho[v(u,w)]^2\,\ud\Omega^2\, ,
\end{equation}
as $v=v(\zeta)=v[\zeta(u,w)]$. For the outgoing radial light geodesic
$u=\mathrm{constant}$, we can define then $\theta_r=\nabla^Ak_A$,
where $k_A=-\partial_A u$ is the normal to the surface. We find that
$\theta_r$ can be written as
\begin{equation}
  \label{eq:wm4}
  \theta_r=\frac{1}{4\sqrt{A\,g_{vv}}}\left[\frac{3}A\dfrac{\ud A}{\ud v}
+\frac{4}\rho\dfrac{\ud \rho}{\ud v}\right] ,
\end{equation}
so that, giving the expressions of $A(v)$ and $\rho(v)$, one can check
that in $\vs<v<\vinf$, one has
\begin{equation}
\theta_\ur > 0\,.
\end{equation}
In particular, this implies that there are no trapped surfaces for
$v\leq v_h$, so that the surface $v_h$ is not an
horizon~\cite{Booth:2005qc}.

In fact, as the $\rho(v)$ behaviour suggests, such the spacetime is of
wormhole kind, the throat of which is precisely located at
$v=\vh$~\cite{Morris:1988cz, Roman:1992xj}. This wormhole is actually
traversible since it does not possess any horizon. However, it does
not connect two asymptotically flat regions since there is a
singularity on one side.

In order to understand the physical meaning of $C$, it is instructive
to study the asymptotic limit. In fact, $\rho\to\infty$ at $v\to
\vinf$ ($C>0$) with
\begin{equation}
  \rho(v) \sim \dfrac{1}{\vinf-v}\,,\quad \xi(v) \sim 1 
  + \left(\gamma_0+1\right) \left(1 - \dfrac{v}{\vinf} \right),
\end{equation}
or, in terms of $\rho$, 
\begin{equation}
\label{eq:xiasympt}
  \xi(\rho) \sim 1+ \dfrac{(\gamma_0+1)C^{1/(\gamma_0-1)}}{\rho} +
  \order{\dfrac{1} {\rho^2}} .
\end{equation}
Similarly, expanding Eq.~(\ref{eq:Aofvinf}) in terms of $\rho$, one
gets
\begin{equation}
  A(\rho) \sim 1 - \dfrac{(\gamma_0 + 1)C^{1/(\gamma_0-1)}}{4 \rho} 
+ \order{\dfrac{1} {\rho^2}}.
\end{equation}
Therefore, for an observer at rest at infinity, $C$ is simply encoding
the tension $T$ of a three-brane associated with the Poincar\'e
invariance
\begin{equation}
C = \left(\dfrac{\gamma_0+1}{4 G_7 T} \right)^{1- \gamma_0},
\end{equation}
Notice that $C$ (or $T$) also fixes the size $\rhoh$ of the wormhole
throat
\begin{equation}
C = \dfrac{1}{\gamma_0^{\gamma_0} \rhoh^{1-\gamma_0}}\, .
\end{equation}

\subsubsection{Embedding function}

\begin{figure}
\begin{center}
\includegraphics[width=8.5cm]{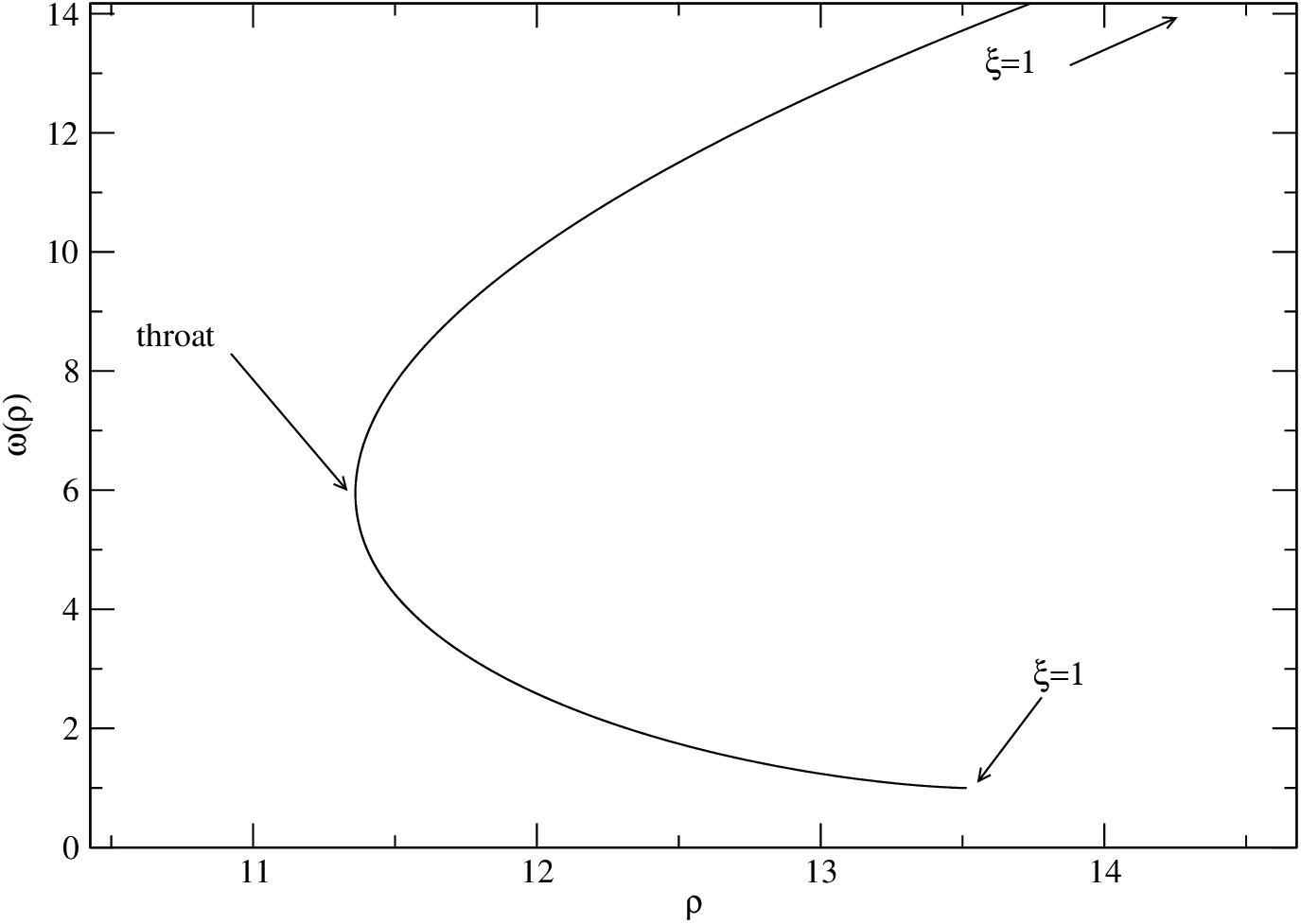}
  \caption{Embedding function $\omega(\rho)$ of the wormhole
    configuration. The plot has been truncated to the $\rho$ values
    for which $\xi \ge 1$ and does not show the singularity.}
\label{fig:whfunc}
\end{center}
\end{figure}

At a fixed four-dimensional location $(t,\vec{x})$, the spatial metric
reads
\begin{equation}
\ud \ell ^2 = \xi(\rho) \ud \rho^2 + \rho^2 \ud \Omega^2.
\end{equation}
Following Ref.~\cite{Morris:1988cz}, this hypersurface can be embedded
in an four-dimensional Euclidian space of metric
\begin{equation}
\ud s_{\ue}^2 = \ud \omega^2 + \ud \rho^2 + \rho^2 \ud \Omega^2.
\end{equation}
This hypersurface has spherically symmetric sections and is described
by the function $\omega(\rho)$ such that
\begin{equation}
\left(\dfrac{\ud \omega}{\ud \rho}\right)^2 = \xi(\rho) - 1,
\end{equation}
The embedding function is no longer defined for $\xi<1$, the
hypersurface being not longer embeddable into an Euclidian space (but
instead it is inside a Minkowski space). In Fig.~\ref{fig:whfunc}, we
have represented $\omega(\rho)$ in the domains for which $\xi \ge
1$. In the asymptotically flat region, from Eq.~(\ref{eq:xiasympt}),
one gets
\begin{equation}
\omega(\rho) \propto \sqrt{\rho}.
\end{equation}

\subsubsection{Radial motion}

It is instructive to study the radial geodesics around the wormhole
solution (see Fig.~\ref{fig:vac:ptc6}). Assuming without loss of
generality that $C=1$, the spacetime is asymptotically flat for $v\to
\vinf=1$. In this case, if $t$ is the proper time of the asymptotic
observer, one has $A\to1$ and thus $A_0=1$ and
\begin{equation}
  A(v) = v^{(\gamma_0+1)/4}.
\end{equation}
The metric for a purely radial motion simplifies into
\begin{equation}
  \ud s^2 = -v^{(\gamma_0+1)/4} \ud t^2 + \dfrac{20
    \gamma_0(\gamma_0-1)^2 v^{\gamma_0-5}}{3
    \left(v^{\gamma_0-1}-1\right)^4} \ud v^2 .
\end{equation}
For a timelike geodesics, denoting by $\tau$ the proper time, a
constant of motion is
\begin{equation}
  A \dfrac{\ud t}{\ud \tau} \equiv \sqrt{A_*} ,
\end{equation}
The meaning of $A_*$ is clear in the asymptotic region: $\ud t/\ud
\tau$ corresponds to the energy per unit mass, so that 
\begin{equation}
  A_*=\dfrac{1}{1-V_\infty^2} \geq 1 ,
\end{equation}
where $V_\infty$ stands for the radial velocity of the particle at
infinity. Conservation of energy along the geodesics now implies
\begin{equation}
  \dfrac{1}{2} \left(\dfrac{\ud v}{\ud \tau} \right) ^2+U_{\mathrm{eff}}(v)=0 ,
\end{equation}
where 
\begin{equation}
  U_{\mathrm{eff}} = \frac1{2g_{vv}}-\frac{A_*}{2g_{vv}A} = -\frac{ 3
    (1 - v^{\gamma_0-1})^4 (A_* - v^{1/4+\gamma_0/4})}{ 40(\gamma_0 -
    1)^2 \gamma_0 v^{5\gamma_0/4-19/4}} ,
\end{equation}
is the effective potential. Since $A_*\geq1$, then
$U_{\rm\mathrm{eff}} \leq 0$ and
$\lim_{v\to0^{+}}U_{\mathrm{eff}}=-\infty$. In this case any radial
geodesics will fall into the singularity. Although the coordinate
$\rho$ goes to infinity as $v$ vanishes, the proper distance $\dsh$
between the singularity and the wormhole throat is actually
finite. {}From Eq.~(\ref{eq:gvv}), one indeed obtains
\begin{equation}
  \dsh = \int_0^{\vinf}\sqrt{g_{vv}}\ud v \simeq 9.988.
\end{equation}
On the other hand one can check that the asymptotically flat region is
at an infinite proper distance from the throat.

The proper time of a free-falling particle along a radial geodesics is
given by
\begin{equation}
  \Delta \tau=-\int_{v_i}^{0} \ud v \sqrt{\frac{g_{vv}A}{A_*-A}}\, .
\label{proprio}
\end{equation}
Since $A_*\geq1$, the geodesics is well defined for all
$0<v<1$. Similarly, the asymptotic time $t$ reads
\begin{equation}
\Delta t=-\int_{v_i}^{0} \ud v \sqrt{\frac{g_{vv} A_*}{A(A_*-A)}}\, ,
\end{equation}
As an example, considering $A_*=2$ and $v_i=0.9$, one finds that the
singularity $v=0$ is reached in the finite proper time $\Delta\tau
\simeq 12.1$, but also for a finite $\Delta t\simeq 43$. For any
$v_i<1$, it takes a finite amount of proper time $\Delta \tau$ and of
$\Delta t$ (the time measured from an observer in the flat region) to
hit the singularity.

Having clarified the coordinate system artifacts for the vacuum case,
we derive in the next section exact solutions for the generic charged
case. In fact, the situation is qualitatively similar and one has to
introduce a new regular coordinate system to fully describe the
manifold.

\section{Charged solutions}
\label{sec:charged}

In this section we derive the charged solutions generated by the
monopole. Instead of discussing the solutions in terms of the radial
coordinate $\rho$, we directly introduce a generalised version of our
new radial coordinate $v$, in a way similar to the vacuum case.

\subsection{First Branch}

As in Sect.~\ref{sec:vacuum}, this branch refers to as choosing a plus
sign for the square root in Eq.~(\ref{dixi}), and we define $v$ such
that

\begin{widetext}
\begin{equation}
\label{eq:xicharged}
\begin{aligned}
  \xi(\rho) & = \dfrac{5\rho^2}{3(\rho^2-1)} \left( \left\{
      \frac{\left[4 \rho^2 + \left(4+\sqrt{10}\right) \rho +
          \sqrt{10}\right] v-(8+2 \sqrt{10}) \rho}{\left[\sqrt{10}
          \rho^2 +\left(4+\sqrt{10} \right) \rho + 4\right] v -(8+2
        \sqrt{10}) \rho}\right\}^2 -1 \right).
\end{aligned}
\end{equation}
The coordinate $v$ is now obtained in terms of $\rho$ by using
Eq.~(\ref{dixi}) and one finds an Abel differential equation of the
second kind, namely
\begin{equation}
  \dfrac{\ud v}{\ud \rho} = \frac{2 \left(4-\sqrt{10}\right) (v-1) v}
  {2 \left(4+\sqrt{10}\right) \rho-(\rho+1) \left(\sqrt{10}
      \rho+4\right) v} \,.
\end{equation}
As for the vacuum case, we can invert the previous relation such that
now $\rho$ is a function of $v$ and one gets
\begin{equation}
  \label{eq:rhovcharg}
  \dfrac{\ud \rho}{\ud v} = -\frac {\sqrt{10} \rho^2}
  {2\left(4 - \sqrt{10}\right)
    (v-1)}
  -\frac{\left(4+\sqrt{10}\right) (v-2) \rho}{2
    \left(4-\sqrt{10}\right) (v-1)v}
  -\frac{2}{\left(4-\sqrt{10}\right) (v-1)} \,.
\end{equation}
This equation is again a Riccati differential equation which can be
linearized by defining $s(v)$ such that
\begin{equation}
\label{eq:drhocharged}
  \rho(v) = \left(\dfrac{4}{5} \sqrt{10} 
    - 2 \right) \dfrac{v-1}{s} \dfrac{\ud s}{\ud v} \,.
\end{equation}
One can use Eq.~(\ref{eq:rhovcharg}) to finally get
\begin{equation}
\label{eq:dscharged}
  \dfrac{ \ud^2 s}{\ud v^2} + \frac{\left(19+4 \sqrt{10}\right) v - 8
    \sqrt{10}-26}{6 (v-1) v} \dfrac{\ud s}{\ud v} 
  + \frac{40+13 \sqrt{10} } {18 (v-1)^2} s 
  =0.
\end{equation}
\end{widetext}
As in the vacuum case, there is still a symmetry in this differential
equation, that is $v\to v/(v-1)$. As a result, each solution in the
range $0<v<1$ can be mapped to the region $v<0$, each solution in the
range $1<v<2$ to the region $v>2$, and conversely. Therefore it is
sufficient to study the interval $0<v<2$. Along this range, the
general solutions of Eqs.~(\ref{eq:drhocharged}) and
(\ref{eq:dscharged}) are
\begin{itemize}
\item $0<v<1$. The function $s(v)$ is given by
\begin{align}
s &= c_1s_1+c_2s_2 ,
\end{align}
where $s_1$ and $s_2$ stand for
\begin{equation}
\label{eq:s12vlt1}
\begin{aligned}
s_1&=\left[\frac{1-v}{(\sqrt{1-v}-1)^4}\right]^{\gamma_1}, \\
s_2&=\left[\frac{1-v}{(\sqrt{1-v}+1)^4}\right]^{\gamma_1} ,
\end{aligned}
\end{equation}
and 
\begin{equation}
\gamma_1 = \dfrac56+\dfrac13\sqrt{10}\,.
\end{equation}
Hence, we obtain $\rho(v)$ as
\begin{equation}
  \rho=\left(\dfrac45\sqrt{10}-2 \right) (v-1)  \dfrac{\dfrac{\ud
      s_1}{\ud v} -
    C \dfrac{\ud s_2}{\ud v}} {s_1-C s_2}  \,,
\label{eq:rhocharged}
\end{equation}
where the integration constant $C=-c_2/c_1$. {}From the above-mentioned
symmetry $\rho(v,C)=-\rho[v/(v-1),C]$, and
$\xi(v,C)=\xi[v/(v-1),C]$. Using this property, it is possible to find
the solution for $v<0$ from the one in $0<v<1$.

\item $ 1 < v < 2$. For this region, one can first extend the previous
  solution to the complex domain, then, by choosing real linear
  combinations of the complex mode functions, one finds
\begin{align}
s &= c_1s_1+c_2s_2 
\end{align}
where $s_1$ and $s_2$ are now given by
\begin{equation}
\label{eq:s12vgt1}
\begin{aligned}
  s_1 & = \left [\frac{v-1} {v^2}\right]^{\gamma_1} \cos[\Ang(v)] ,\\
  s_2 & = \left [\frac{v-1} {v^2}\right]^{\gamma_1} \sin[\Ang(v)],
\end{aligned}
\end{equation}
with
\begin{equation}
  \label{eq:DEL}
  \Ang(v)\equiv\gamma_1 \left[\pi - 4 \arccos\left( \sqrt{1-\dfrac{1}{v}}\right)
    \right] .
\end{equation}
Notice that $\Ang(2)=0$ so that $s_{1}$ and $s_2$ are, respectively,
even and odd under the symmetry $v \to v/(v-1)$. The solution for
$\rho(v)$ is still given by Eq.~(\ref{eq:rhocharged}) in terms of the
above $s_1$ and $s_2$ functions and therefore explicitly exhibits the
above-mentioned symmetry. Because of the antisymmetry of $s_2$, we
now have $\rho(v,C)=-\rho[v/(v-1),-C]$, and
$\xi(v,C)=\xi[v/(v-1),-C]$.
\end{itemize}

\begin{widetext}

\subsection{Second branch}

This branch corresponds to the plus sign in Eq.~(\ref{dixi}) and, this
time, we define $v(\rho)$ such that

\begin{equation}
  \xi(\rho) = \frac{5\rho^2}{3(\rho^2-1)} \left( \left\{
      \frac{\left[-4 \rho^2 +
          \left(4 + \sqrt{10}\right) \rho - \sqrt{10} \right]
        v - \left (8 + 2 \sqrt{10}\right) \rho} { \left[ \sqrt{10}
          \rho^2 - \left( 4 + \sqrt{10} \right) \rho + 4 \right] v
        + \left( 8 + 2 \sqrt{10} \right) \rho}
    \right\}^2 - 1 \right).
\end{equation}
The Abel equation for $v(\rho)$ reads
\begin{equation}
  \dfrac{\ud v}{\ud \rho} = \frac{2 \left(4-\sqrt{10}\right) (v-1) v}
  {2 \left(4+\sqrt{10}\right) \rho+(\rho-1) \left(\sqrt{10}
      \rho-4\right) v} \,,
\end{equation}
which, as for the first branch, can be written as a Riccati equation,
and one gets
\begin{equation}
  \dfrac{\ud \rho}{\ud v} = \frac{\sqrt{10}\rho^2}
  {2\left(4-\sqrt{10}\right) 
    (v-1)}
  -\frac{\left(4+\sqrt{10}\right)\left(v-2\right) \rho} 
  {2 \left(4-\sqrt{10}\right) (v-1)v}
  +\frac{2}{\left(4-\sqrt{10}\right) (v-1)} \,.
\end{equation}
\end{widetext}
Comparing this expression to Eq.~(\ref{eq:rhovcharg}) shows that the
solutions $\rho_2(v)=-\rho_1(v)$, where $\rho_1(v)$ denotes the
solution of the first branch. As for the vacuum case, for every
$v\neq0$ and $v\ne1$, there is always a positive solution for $\rho$:
if $\rho$ is negative for a branch, it is positive in the other and
vice-versa. We also have $\xi_2(v)=\xi_1(v)$ in between the two
branches.

As for the vacuum case, we join the two branches by defining
$\rho(v)\equiv |\rho_1(v)| = |\rho_2(v)|$ and $\xi(v)\equiv
\xi_1(v)=\xi_2(v)$ for all $0<v<2$. In the following, we discuss the
nature of the solution over the coordinate $v$.

\subsection{Singularities}

At the boundary of the intervals $0<v<1$ and $1<v<2$, the metric
coefficient and coordinates behaves as
\begin{align}
  \lim_{v\to0^{+}}\rho & = +\infty , \qquad \lim_{v\to0^{+}}\xi=0, \\
  \qquad \lim_{v\to0^{+}}K  &= +\infty , \qquad\lim_{v\to0^{+}}R  = 0 ,
\end{align}
and a singularity is present for $v=0$ (and $\rho \to \infty$). As
$v\to1$ we have
\begin{equation}
\lim_{v\to1}K  <\infty , \qquad \lim_{v\to1^{\pm}}\rho  = 1 ,\\
\end{equation}
and one may wonder if the manifold patches can be matched at
$v=1$. However, the metric factor $\xi(v)$ is not continuous at that
point and one has
\begin{align}
  \lim_{v\to1^{-}}\xi&=\frac{160 C}{\left(13-6\gamma_1\right)^2
    (C+1)^2} \, ,\\
  \lim_{v\to1^{+}}\xi&=\frac{40 \left(D^2+1\right)\,(13-6
    \gamma_1)^{-2}} { [\sin (\pi \gamma_1)-D \cos(\pi \gamma_1)]^2}\,,
\end{align}
where we have allowed a different integration constant $D$ for the
regions $v>1$. In fact, there is no real solution for the constant $D$
to match the value of $\xi$ for all $C$. As a result, these two
patches cannot be prolonged one onto another.

For $v \to 2$, and assuming $C\neq 0$, there is no singularity and one
has
\begin{equation}
  \lim_{v\to 2}\rho  =|C|  ,
  \qquad \lim_{v\to 2}\xi=\frac{40 C^2(C^2+1)}{[C^2 (6\gamma_1-5)+8]^2} \,.
\end{equation}

For the intermediate regions, as for the vacuum case, the metric in
the $\rho$ coordinate system possesses a null surface, which ends up
being the throat of a wormhole, if there is a value
$\vh$ for which
\begin{equation}
\lim_{v\to\vh} \xi = \infty.
\end{equation}
A singularity is present when the Kretschmann scalar diverges, i.e.,
for if for the values $\vs$ such that
\begin{equation}
\lim_{v \to \vs} \xi = 0.
\end{equation}
We will also separate the patches explored by the $\rho$ coordinates
according to the values $\vinf$ for which
\begin{equation}
\lim_{v \to \vinf} \rho = \infty.
\end{equation}
The values of $\vh$, $\vs$ and $\vinf$ are obtained by taking the
corresponding limits in Eqs.~(\ref{eq:xicharged}) and
(\ref{eq:rhocharged}). In general, these equations do not have any
obvious analytical solution and they have been numerically solved as a
function of $C$ on the two domains $0<v<1$ and $1<v<2$.

\subsubsection{Over $0<v<1$}

Taking the limit $\rho\to\infty$ in Eq.~(\ref{eq:rhocharged}) gives
\begin{equation}
  \label{eq:Crinf01}
  \vinf=1-\left[\frac{1-C^{1/(4\gamma_1)}}{1+C^{1/(4\gamma_1)}}\right],
\end{equation}
which is such that $0<\vinf<1$ for $0<C<1$ only.

{}From Eq.~(\ref{eq:xicharged}), the null surface position is solution of
\begin{equation}
\hfunc^{(1)}(\vh,C)=0,
\end{equation}
where $\hfunc^{(1)}(v,C)$ is given by Eq.~(\ref{eq:rhor01}). Finally,
except for some particular cases, the singularity coincides with the
points where $\rho(v)$ vanishes, so that $\vs$ is solution of
\begin{equation}
\sfunc^{(1)}(\vs,C)=0,
\end{equation}
$\sfunc^{(1)}(v,C)$ being given in Eq.~(\ref{eq:csing01}).

\begin{figure}
\begin{center}
\includegraphics[width=8.5cm]{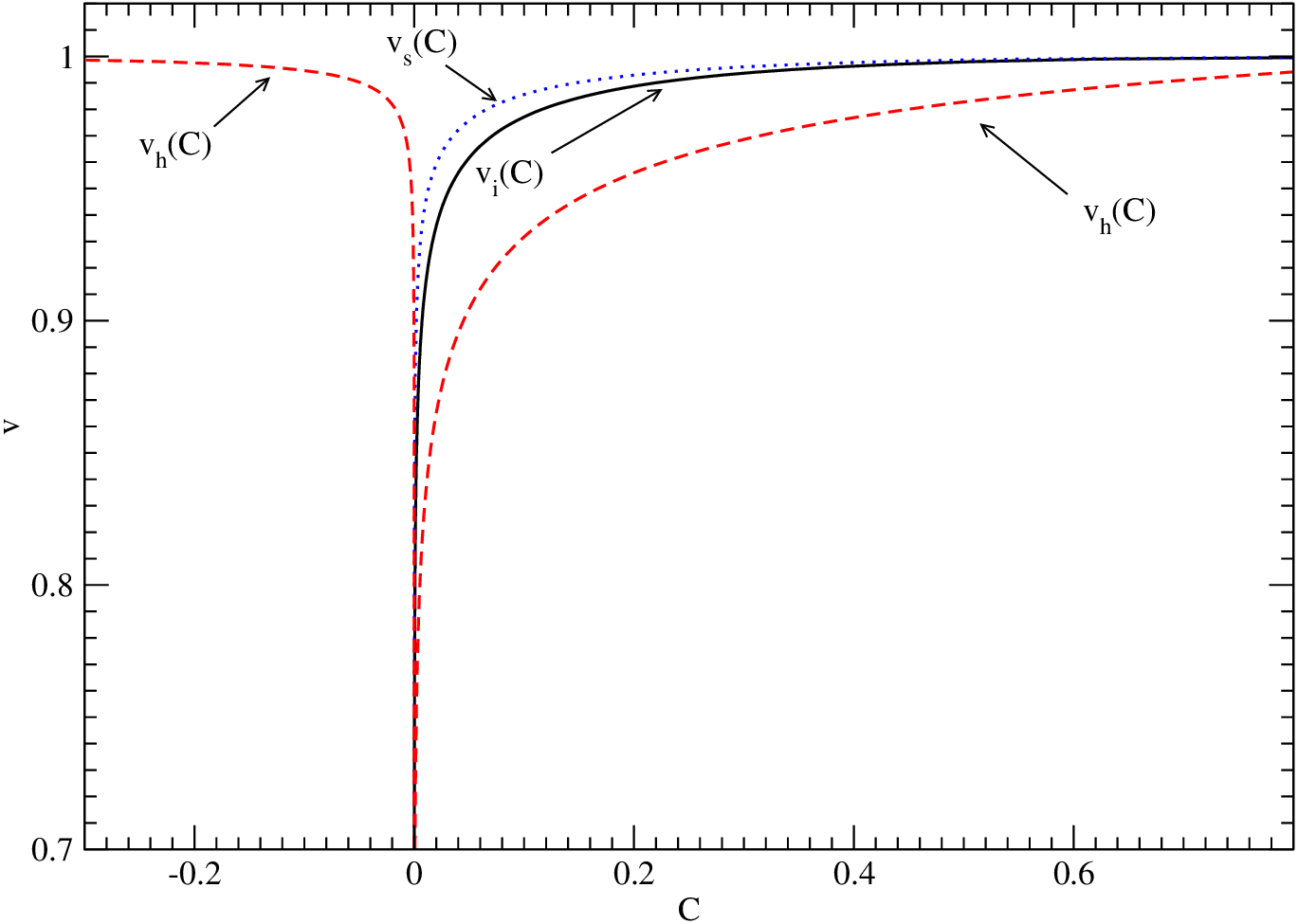}
\caption{Location of the singular points and asymptotic
  $\rho$-domains as functions of $C$ in the region $0<v<1$. For
  $v\to0$, $\xi$ vanishes and a singularity appears (not represented,
  see text).}
\label{fig:V01:cs}
\end{center}
\end{figure}

The solutions of the above equations have been plotted in
Fig.~\ref{fig:V01:cs} where $\vh$, $\vs$ and $\vinf$ are functions of
$C$. {}From this plot, one can distinguish three cases:
\begin{itemize}

\item For $C<0$, we have a singularity for $v\to0$, a null surface (for
  $-1<C<0$) and $\xi<0$ for all $v$ in $0<v<1$. We will therefore not
  consider any longer this pathological case.

\item For $0<C<1$ and $0<v<\vinf$, we have a wormhole configuration
  whose solution is shown in Fig.~\ref{fig:V01:c2}. This case is
  discussed in more details in the next section.

\item For $0<C<1$, and $\vinf<v<1$, $\xi$ vanishes without changing
  its sign and a naked singularity appears. Notice that the proper
  distance in this interval tends to infinity, although $\rho\to1$ as
  $v\to 1^{-}$.

\item $C>1$. In this case there is a naked singularity in $\vs=0$.

\end{itemize}

In the next section, we perform the same analysis on the domain
$0<v<2$.

\subsubsection{Over $1<v<2$}

It is interesting to notice the difference with respect to the
previous case, as now $\xi$ is positive definite. The existence of
coordinate singularities is still given by the condition
$\xi\to\infty$ and $\vh$ is solution of
\begin{equation}
\hfunc^{(2)}(\vh,C)=0,
\end{equation}
where $\hfunc^{(2)}(v,C)$ is derived in the Appendix. In the domain
$1<v<2$, Eq.~(\ref{eq:hfunc12}) is well defined only for
\begin{equation}
  1<v<\frac{-(6 + 24 \gamma_1)+4 \sqrt{78 \gamma_1+15}}{12 \gamma_1-23}
  \simeq 1.003 .
\end{equation}
The singularities are given by the values for $v$ at which $\xi$
vanishes and are the zeros of
\begin{equation}
\sfunc^{(2)}(\vs,C) = 0,
\end{equation}
where $\sfunc^{(2)}(v,C)$ is given by Eq.~(\ref{eq:sfunc12}).

The conditions for the points $\vinf$ at which $\rho\to\infty$, can be
simplified into
\begin{equation}
  \label{eq:rhoinfCH}
  \vinf= \dfrac{1}{1-\cos^2
    \left[\dfrac\pi4-\dfrac{\arctan(C^{-1})}{4\gamma_1} \right]}\, .
\end{equation}

\begin{figure}[t]
\begin{center}
\includegraphics[width=8.7cm]{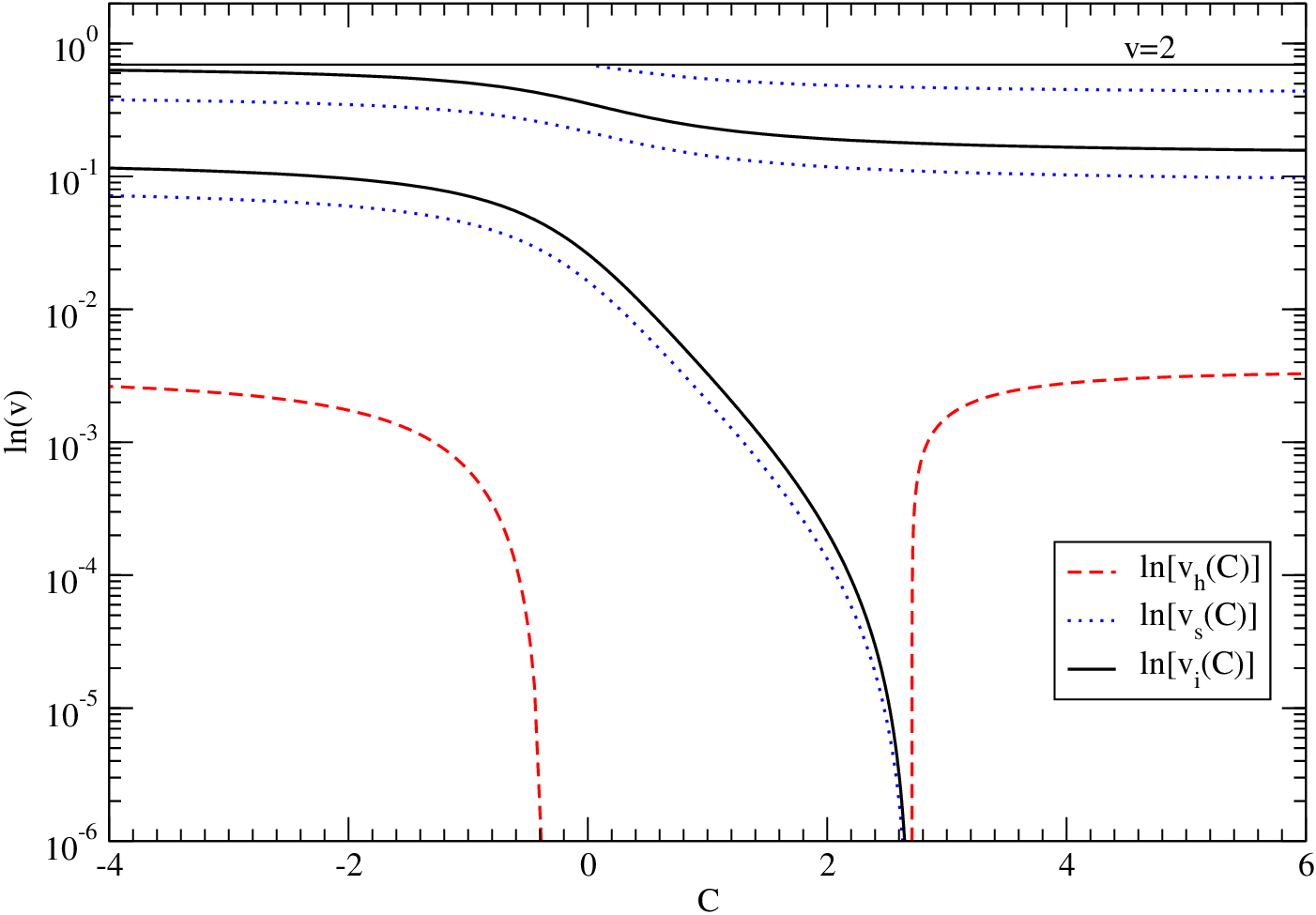}
\caption{Position of the singularities and asymptotic
  $\rho$-domains as a function of $C$ in the region $1<v<2$. These
  patches exhibit naked singularities.}
\label{fig:Cvgt1}
\end{center}
\end{figure}

The condition of existence and dependency of $\vh$, $\vs$ and $\vinf$
with respect to the integration constant $C$ have been represented in
Fig.~\ref{fig:Cvgt1}.

For a given value of $C$, this plot shows whether and where there are
singularities, bounded or unbounded $\rho$-domains and the null
surfaces. Among the many possibilities, none of them seems to
represent physically interesting regions, as there are naked
singularities and no horizon connected to asymptotically flat
space-time. Therefore we will not consider further this part of the
full solution, although some of these configurations may have some
interest once regularised.

In the next section, we study in more details the charged wormhole
solution exhibited for $0<v<1$.

\subsection{Charged wormhole configuration }

\begin{figure}
\begin{center}
\includegraphics[width=8.5cm]{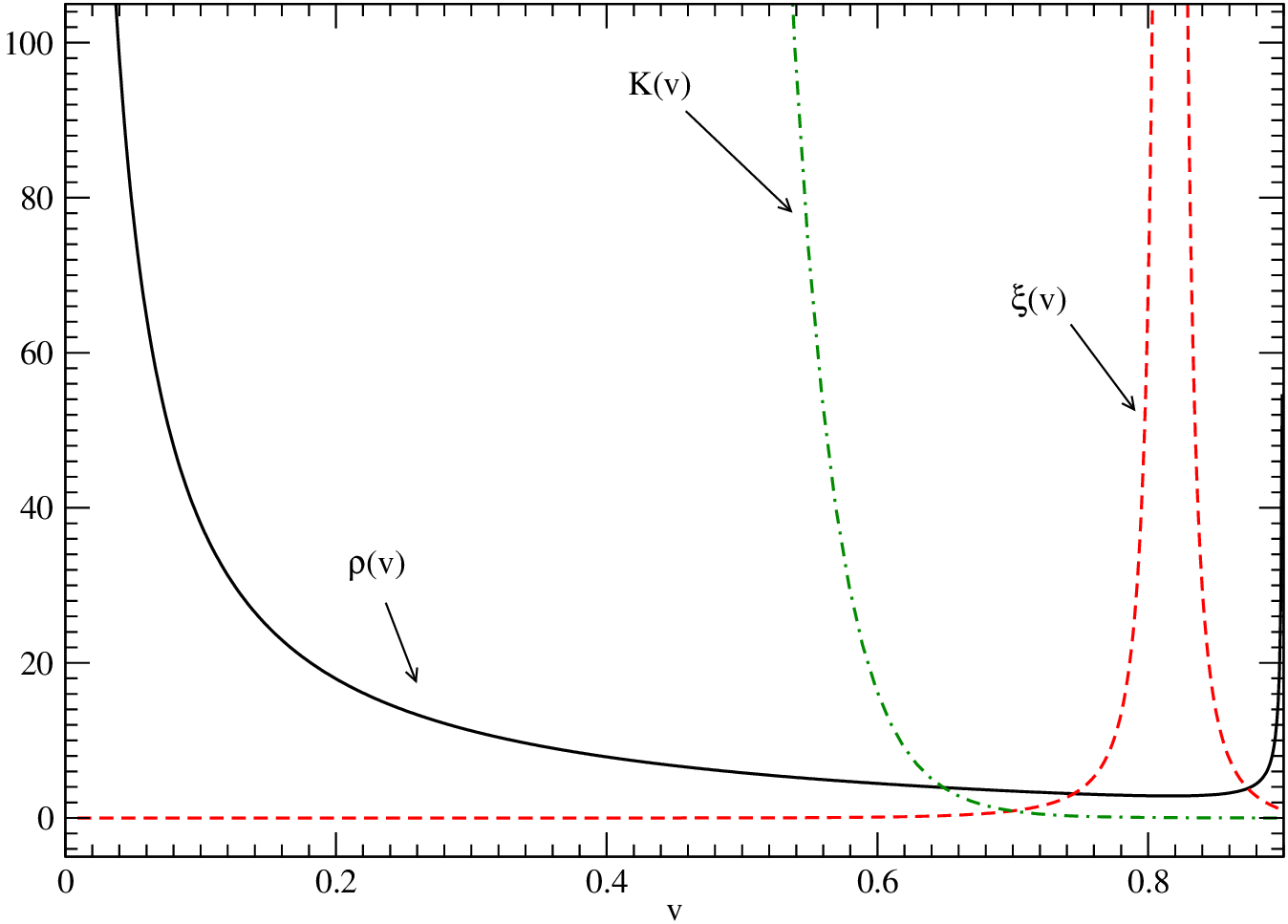}
\caption{Charged wormhole solution. There is a singularity for
  $v\to0$, the wormhole throat is located in $v=\vh<\vinf$, at finite
  proper distance from the singularity, and the space is
  asymptotically flat as $v\to\vinf$. For convenience, the brane
  tension has been chosen such that $\vinf=9/10$.}
\label{fig:V01:c2}
\end{center}
\end{figure}

We denote by this charged wormhole configuration, the solution
obtained in the region $0<v<\vinf$. By definition
\begin{equation}
\lim_{v\to    \vinf^{-}}\rho(v)=+\infty.
\end{equation}
Defining the following polynomials
\begin{align}
P_1(\rho)&=4 \rho^2+(\sqrt{10}+4)\rho+\sqrt{10},\\
P_2(\rho)&=\sqrt{10}\rho^2+(\sqrt{10}+4)\rho+4,\\
P_3(\rho)&=2(\sqrt{10}+4)\rho,
\end{align}
one gets from Eq.~(\ref{eq:xicharged})
\begin{equation}
  \xi(v)=\frac{10\rho^2 v}{(P_2v-P_3)^2}[(\rho+1)^2v-4\rho] ,
\end{equation}
and from Eq.~(\ref{eq:rhovcharg})
\begin{equation}
  \dfrac{\ud \rho}{\ud v}=\frac{P_2 v-P_3}{2(4-\sqrt{10})v(1-v)} .
\end{equation}
Therefore in the $v$ variable, the metric component $g_{vv}$ becomes
\begin{equation}
  g_{vv}=\xi \left(\dfrac{\ud \rho}{\ud v} \right)^2 =
  \dfrac{5} {2 (4-\sqrt{10})^2} \dfrac{\rho^2[(\rho+1)^2v-4\rho]}{v(1-v)^2}\, ,
\end{equation}
where $\rho(v)$ is explicitly given by Eq.~(\ref{eq:rhocharged}). It
is already clear, using the $v$ variable that the $g_{vv}$ component
of the metric is not any longer divergent at the $\xi(\rho)$ null
surface. As for the uncharged case studied in Sect.~\ref{sec:vacuum},
this configuration is in fact a wormhole which does not possess any
horizon. The proper distance from the singularity ($\vs=0$) to the
throat is actually finite and this configuration is reminiscent with
the uncharged one:
\begin{equation}
  \Delta s=\int_{0}^{\vh}\sqrt{g_{vv}} \ud v <\infty.
\end{equation}
Also in this case, $v$ grows continuously from the singularity point
to an asymptotically flat region whereas $\rho$ shows a multivalued
typical behaviour of a wormhole configuration. The various geometrical
quantities associated with this solution have been represented in
Fig.~\ref{fig:V01:c2}.

\section{Conclusions}

We have presented exact solutions of the Einstein equations in seven
dimensions in the presence of matter in the form of a monopolar
magnetic field. The solutions have been derived for compactified
spacetimes of the form $M^4\times R\times S^2$ for which the
extra-dimensions exhibit spherical symmetry. The form field considered
is generated by a spatially extended Dirac monopole present in the
origin and found to generate a charged black brane configuration. This
charged black brane will be in general a singularity of the spacetime
and may or not be naked. In the regular case, a null surface is
situated at a finite proper distance from the brane and 
corresponds to the throat of a wormhole configuration. In order to
derive such an explicit solution that covers the whole spacetime, we
have proposed a new coordinate system which cures the incompleteness
of the natural spherical coordinates ($\rho$ coordinate). Such a
choice of coordinates clarify the issues of the apparent ``holes'' in
the manifold and our approach may be applied to other similar
braneworld models. Notice that the solution described here correspond
to the asymptotic behaviour of more complicated models exhibiting a
residual $U(1)$ symmetry, as the ones described in
Refs.~\cite{Ringeval:2004ju,DeFelice:2008af}. Let us also notice that
we have not discussed the stability of our solutions. Since they
correspond to wormhole configurations with spatially unbounded
traverse sections, one may be concerned about instabilities. It is
indeed possible that such systems, in the way presented here, may be
unstable against breakdown into standard seven-dimensional black
holes~\cite{Gregory:1993vy, Gregory:1994bj, Gregory:1994tw}. However,
since such solutions also match with exterior spacetime of
higher-dimensional topological defects, the stability of the later
remains an open question for future works.

\begin{acknowledgments}
  We would like to thank an anonymous referee for his enlightening
  comments. This work is supported by the Belgian
  Federal Office for Science, Technical and Cultural Affairs, under
  the Inter-university Attraction Pole grant P6/11.
\end{acknowledgments}

\begin{appendix}

\begin{widetext}
\section{Geometrical quantities}

\subsection{Scalar invariants}
\label{app:grinv}

{}From the metric of Eq.~(\ref{eq:metric}), the Ricci and Kretschmann
scalars are given by
\begin{align}
\label{eq:Rscalar}
  R&=-\frac{4 A''}{\xi A}
  +\frac{2A'\xi'}{A\xi^2}
  -\frac{8A'}{r\xi A}
  +\frac{2\xi'}{r\xi^2}
  +\frac2{r^2}
  -\frac2{r^2\xi}
  -\frac{{A'}^2}{\xi A^2} ,\\
  K&=-\frac{A''{A'}^2}{\xi^2A^3}+\dfrac14\frac{{A'}^2{\xi'}^2}{\xi^4A^2}
  +\dfrac12\frac{{A'}^3\xi'}{\xi^3A^3}
  +\frac{{A''}^2}{\xi^2A^2}
  +\frac{2{A'}^2}{\rho^2\xi^2A^2} 
   + \frac1{\rho^4} +\dfrac12\frac{{\xi'}^2}{\rho^2\xi^4}-\frac{A''A'\xi'}{\xi^3A^2}
  -\frac2{\rho^4\xi}+\dfrac58\frac{{A'}^4}{\xi^2A^4}
  +\frac1{\rho^4\xi^2} .
\label{eq:Kscalar}
\end{align}

\subsection{Singularities}

As discussed in Sect.~\ref{sec:charged}, the location of singularities
in the $\rho$-coordinate system for the charged solution are given by
respectively considering the limit $\xi \to \infty$ and $\xi \to 0$ in
Eq.~(\ref{eq:xicharged}).

\subsubsection{Region $0<v<1$}

For $0<v<1$, the null surface location $\vh$ is given by the root of the
limit $\xi(v)\to\infty$ in Eq.~(\ref{eq:xicharged}) where $\rho(v)$ is
obtained from Eq.~(\ref{eq:rhocharged}) with the mode functions $s_1$
and $s_2$ of Eq.~(\ref{eq:s12vlt1}). After some calculations, $\vh$ is
found to be the root of
\begin{align}
&  \hfunc^{(1)}(v,C) = -C +
    \dfrac{2 \left(\sqrt{1-v}+1\right)^{6-8 \gamma_1} v^{4 \gamma_1}}{
      6 \gamma_1-13 }
    \nonumber \\ & \times \frac{6 \gamma_1 (v-1)\pm\sqrt{2}
        \sqrt{(v-1) \{12 \gamma_1 [v (v+4)-4]+(12-23 v) v-12\}}+3
        v-3 }%
    {v \left\{v\left[\left(\sqrt{1-v}+8\right) v^2-8 \left(4
            \sqrt{1-v}+11\right) v+160 \sqrt{1-v}+272\right]-64
        \left(4 \sqrt{1-v}+5\right)\right\}+128
      \left(\sqrt{1-v}+1\right)} \, .
  \label{eq:rhor01}
\end{align}
Similarly, the singularities location $\vs$ are given by the root of
\begin{equation}
  \label{eq:csing01}
  \sfunc^{(1)}(v,C) = -C +   \frac{ v+2 \sqrt{1-v}-2} {v-2
    \sqrt{1- v}-2}
  \left[\frac{\left(\sqrt{1-v}-1\right)^4}{v^2-8 v+4 (1-v)^{3/2}+4
      \sqrt{1-v}+8}\right]^{\gamma_1} \,.
\end{equation}

\subsubsection{Region $1<v<2$}

As for the previous domain, one obtains $\vh$ and $\vs$ from the
corresponding limits of Eq.~(\ref{eq:xicharged}), with $\rho(v)$ given
by Eq.~(\ref{eq:rhocharged}) but with now the mode functions $s_1$ and
$s_2$ of Eq.~(\ref{eq:s12vgt1}). The locations $\vh$ end up being the
root of
\begin{align}
  & \hfunc^{(2)}(v,C) =-C \nonumber \\ & + \dfrac{-6 (2 \gamma_1+1)
    \sqrt{v-1}-2 (6 \gamma_1-13) \sqrt{v-1} \cos
    [2\Ang(v)]+(6\gamma_1-13) (v-2) \sin[2 \Ang(v)]} {(6 \gamma_1-13)
    (v-2) \cos[2 \Ang(v)]\pm 2 \sqrt{-24 \gamma_1\left(v^2+4
        v-4\right)+46 v^2-24 v+24}+2(6 \gamma_1-13)
    \sqrt{v-1}\sin[2\Ang(v)]}.
\label{eq:hfunc12}
\end{align}
The singularities positions $\vs$ are obtained by the zeros of
\begin{equation}
  \sfunc^{(2)}(v,C) = C + \frac{(v-2) \sqrt{v-1} \cos[\Ang(v)]
    +2 (v-1) \sin[\Ang(v)]}%
  {2 (v-1) \cos[\Ang(v)]-(v-2) \sqrt{v-1} \sin[\Ang(v)]} .
\label{eq:sfunc12}
\end{equation}
In both of the previous expression, the function $\Ang(v)$ is given by
Eq.~(\ref{eq:DEL}).

\end{widetext}

\end{appendix}

\bibliography{bibpole}

\end{document}